\documentstyle [12pt] {article}
\documentstyle {mystyle}

\newcommand{\nummer}[1]{\hskip 12 true cm #1 \par}
\newcommand{\monat}[1]{\vspace{-14pt}\hskip 12 true cm #1
                       \par \vspace*{1 cm}}
\newcommand{\titel}[1]{{\renewcommand{\thefootnote}{\fnsymbol{footnote}}
                       \Large\bf\vskip 0 true cm
                       \begin{center}#1\end{center}
                       \setcounter{footnote}{0}}
                       \normalsize\vskip 1.2 true cm}
\newcommand{\autor}[1]{{
                       \renewcommand{\thefootnote}{\fnsymbol{footnote}}
                       \begin{center} {\large #1 }\end{center}}
                       \setcounter{footnote}{0}}
\newcommand{\adresse}[1]{\vspace*{-1.1 true cm}\begin{center} {\it #1 }
                         \end{center}
                         \vskip 0.5cm}

\font\dickmi=cmbx10 scaled\magstep2

\newcommand{\bye}{\end{document}}
\newcommand{\be}{\begin{equation}}
\newcommand{\ee}{\end{equation}}
\newcommand{\bes}{\begin{eqnarray}}
\newcommand{\ees}{\end{eqnarray}}
\newcommand{\ema}{\end {array} \right)}
\newcommand{\pslash}{\kern 0.1 em p\kern -0.45em /}
\newcommand{\dslash}{\kern 0.1 em \partial\kern -0.55em /}
\newcommand{\sla}[1]{\kern 0.2 em #1\kern -0.55em /}

\newcommand{\LRA}{\Leftrightarrow}

\newcommand{\R}{\mbox{I\kern -0.22em R\kern 0.30em}}
\newcommand{\N}{\mbox{I\kern -0.22em N\kern 0.30em}}
\newcommand{\C}{\mbox{\kern 0.20em \raisebox{0.09ex}{\rule{0.08ex}{1.22ex}}
                \kern -0.60em C\kern 0.30em}}
\newcommand{\Z}{\mbox{\sf Z\kern -0.40em Z\kern 0.30em}}

\newcommand{\Hext}{{\cal H}_{ext}}

\newcommand{\Hz}[1]{{\cal H}_{#1}}

\newcommand{\noSl}{\kern 0.20em :\kern -0.10em :}
\newcommand{\noSr}{: \kern -0.10em :\kern 0.30em}
\newcommand{\aYM}{\alpha_{Y\! M}}
\newcommand{\kint}[1]{\oint\!\!{d{#1}\over 2\pi i}\;}

\newcommand{\cb}{\overline{c}}

\newcommand{\etb}{\overline{\eta}}

\newcommand{\gh}[1]{{\it gh}(#1)}
\newcommand{\eqv}{\Leftrightarrow}
\newcommand{\Om}{{\bf \Omega}}
\newcommand{\Omb}{{\bf \overline{\Omega}}}

\newcommand{\bra}{<\!\!\Psi |}
\newcommand{\ket}{|\Psi\!\!>\:}

\newcommand{\ketp}{|\Psi^\prime\!>\:}
\newcommand{\bral}{<\!\!\Lambda |}
\newcommand{\ketl}{|\Lambda\!\!>\:}

\newcommand{\ketph}{|\Phi\!\!>\:}
\newcommand{\vacr}{|0\!\!>\:}
\newcommand{\vacl}{<\!\!0|}
\newcommand{\hws}[1]{| {\bf #1} \!\!>}

\begin{document}

\begin{titlepage}
\nummer{NIKHEF-H/92-13}
\monat{September 9, 1992}
%\monat{{\it(revised version)}}
{\vskip 1cm}
\titel{BRST Quantization of 2d-Yang Mills Theory with Anomalies}
\autor{Wolfgang Kalau\footnote{bitnet adr.:
t11@nikhefh.nikhef.nl}\footnote{Address after Oct.~1,  1992:
Institut f\"ur Physik, Johannes-Gutenberg-Universit\"at Mainz, Staudinger
Weg 7, D-6500 Mainz, Germany}}
{\vskip 0.5cm}
\adresse{NIKHEF-H\\
P.O.Box 41882\\
NL-1009 DB Amsterdam}
{\vskip 2cm}
\begin{abstract}
In this paper we discuss the BRST-quantization of anomalous 2d-Yang Mills
(YM) theory. Since we use an oscillator basis for the YM-Fock-space the anomaly
appears already for a pure YM-system and the constraints form a Kac-Moody
algebra with negative central charge. We also discuss the coupling of
chiral fermions and find that the BRST-cohomology for systems with chiral
fermions in a sufficiently large representation of the gauge group is
completely equivalent to the cohomology of the finite dimensional gauge
group. For pure YM theory or YM theory coupled to chiral fermions in  small
representations there exists an infinite number of inequivalent cohomology
classes.  This is discussed in some detail for the example of $SU(2)$.
\end{abstract}
\end{titlepage}

\section{Introduction}
It has turned out that the BRST quantization method is a quite general
and useful method to construct quantum theories for systems with first-class
constraints. However, there are systems for which some of the first-class
constraints at the classical level turn out to be second-class constraints
at the quantum level. Such systems are known as anomalous theories and for
those the usual BRST procedure fails.

Since the discovery of anomalies
\cite{ABJ} a lot of work has been performed to unravel their general structure
and to construct consistent quantum theories for those systems. In this context
Electrodynamics (Schwinger model) and Yang-Mills theory coupled to chiral
fermions in two dimensions have served as instructive and interesting toy
models. For recent work on this topic see e.g. \cite{JMi,MbP,HeHo,jurek} and
references therein.

The study of 2 dimensional Yang-Mills (YM) theories is closely related to the
study  of Kac-Moody algebras since the constraint functions for the phase space
of YM theory on a circle are maps from $S^1$ to the algebra of the gauge group.
The anomaly appears at the quantum level as a central charge for the Kac-Moody
algebra. There is a natural complex structure for Kac-Moody algebras and
therefore the generalized BRST formalism developed in \cite{JW,JWW,wk}
can be applied to this system which will lead to a complex non-Hermitian BRST
operator. The BRST-quantization of Kac-Moody algebra has been studied by
several authors \cite{BiGe,Ka,BoHl}. However, they have constructed Hermitian
BRST operators which turned out not to be nilpotent for a non-vanishing
central charge.

This article is organized as follows.
Section 2 contains the canonical analysis of 2d YM-theory with the derivation
of the classical constraint system which is a Kac-Moody algebra without a
central charge. In section 3 we construct a quantum mechanical state space for
this system.
Instead of choosing a coordinate representation for the operators
corresponding to the classical fields we shall quantize the system in an
oscillator basis. As a consequence we find an anomalous term already for the
pure Yang-Mills system. In section 4 we give a brief introduction to the
general BRST formalism for holomorphic second-class constraints which is
applied to 2d YM-theory in section 5.
The BRST-Laplacian and its gauge-fixing properties are disussed in section 6.
In section 7 we analyse the general structure of the cohomology of the
non-Hermitian BRST operator and we will study $SU(2)$-YM-theory as a specific
example in section 8. We end this article with some conclusions in section 9.

\section{Canonical Structure of YM-Theory on $S^1$}
The action for 2 dimensional YM-theory on a circle coupled to chiral fermions
is given by
\bes
S=\int\!\!d^2\!x\; \left( -{1\over 4} F_{\mu\nu}^a F^{\mu\nu}_a - {i\over 2}
{\overline \Psi}^i \sla{D}_{ij}{1\over 2}(1 + \gamma^5)\Psi^j\right) \;\; .
\label{YMac}
\ees
$F_{\mu\nu}^a = \partial_{[\mu}A^a_{\nu ]} + f^a_{bc}A^b_\mu A^c_\nu$ denotes
the field strength of the gauge potential $A^a_\mu$ of some simple gauge
group $G$ with algebra $\cal G$ where the structure constants $f_{abc}$ are
chosen to be completely antisymmetric. The fermions are real Majorana fermions
in some representation of $\cal G$. The generators of the algebra in this
representation are given by ${\rho_a}_i^j$. For the $\gamma$-matrices we choose
the the following convention
$$
\gamma^0 = \left( \begin{array}{cc} 0 & -1\\ 1 &0\end{array}\right)\;\; ,\;\;
\gamma^1 = \left( \begin{array}{cc} 0 & 1\\ 1 &0\end{array}\right)\;\; ,\;\;
\gamma^5 = \left( \begin{array}{cc} -1 & 0\\ 0 &1\end{array}\right)\;\;
$$
and $\sla{D}_{ij} = \dslash\delta_{ij} + \sla{A}^a \rho_{a\; ij}$ is the
covariant derivative. Denoting by $\Psi^i_+ = {1\over 2}(1+\gamma^5)
\Psi^i$ the
right handed component of the spinor, we find for the canonical momenta:
\bes
{ \delta S \over \delta \partial_0 A^a_0(x)} &=& 0\;\;,\label{mo1}\\
{ \delta S \over \delta \partial_0 A^a_1(x)} &=& F^a_{01}(x) =: E^a(x)\;\;,
\label{mo2}\\
{ \delta S \over \delta \partial_0 \Psi^i_+(x)} &=& {i\over2}\Psi^i_+(x)\;\;.
\label{mo3}
\ees
Eq.~(\ref{mo3}) leads to the usual second-class constraint for fermions which
can be eliminated by introducing the Dirac bracket. We therefore have the
standard bracket for the fermions:
\bes
\{\Psi^i_+(x),\Psi^j_+(y)\} = i \delta(x-y)\delta_{ij}
\ees
Eq.~(\ref{mo1}) is a primary constraint and the requirement that it
be a constant in time leads to the
secondary constraint, the Gau\ss{} law:
\bes
G_a(x) = \partial_x E_a(x) + f_{ab}^c A^b(x)E_c(x) -{i\over2}\rho_{a\;ij}
\Psi^i_+(x)\Psi^j_+(x)=0\;\; .
\ees
With this we can now write down the Hamiltonian as
\bes
H = \int_{0}^{2\pi}\!\!dx\; {1\over 2} E^aE_a + {i\over 2}\Psi^i_+D_{x\;ij}
\Psi^j_+
\;\; ,
\ees
where $A_0^a$ has been eliminated since it contains no dynamical degree of
freedom.

In the next step we transform the fields to their momentum representation
which on the circle has the nice property that it is discrete. The Fourier
components of the fields are given by
\bes
A^a_n = \kint{z} z^{n-1}A^a(z)\;\;&,&\;\;
E^a_n = \kint{z} z^{n-1}E^a(z)\;\; ,\;\; n \in \Z\\
u^i_n &=& \kint{z} z^{n-1}\Psi^i_+(z)\;,
\ees
where the integration is taken along the unit circle in the complex plane.
The Fourier index of the fermions is either integer valued for periodic
Ramond (R) fermions or half integer  for antiperiodic Neveu-Schwarz (NS)
fermions. Since all fields are real their Fourier components are restricted
to
\bes
{A_n^a}^\dagger = A_{-n}^a\;\; ,\;\; {E_n^a}^\dagger = E_{-n}^a\;\; ,\;\;
{u_n^i}^\dagger = u_{-n}^i\;\; . \label{clherm}
\ees
The canonical Poisson brackets for these variables are
\bes
\{A^a_n,E^b_m\} = -\delta^{ab}\delta_{n,-m}\;\;, \;\;
\{u^i_n,u^j_m\} = -i\delta^{ij}\delta_{n,-m}\;\;.
\ees
We can now also transform the constraints to the momentum representation
by
\bes
G^n_a= \kint{z} z^{n-1}G_a(z)\;\; n\in\Z
\ees
and rewrite them in terms of Fourier modes of the fields
\bes
G^n_a = inE^n_a + \sum_{r\in \Z}f_{ab}^c A^b_{-r}E^{r+n}_c - \sum_{r\in \Z}
{i\over 2} {\rho_a}_j^i u_i^{n-r-p}u_{r+p}^j \;\; ,
%\left\{\begin{array}{cc}p= 0& R-case\\p0{1\over 2}& NS-case\end{array}\right.
\ees
where $p=0$ in the R-case and $p={1\over 2}$ for the NS-case.
In these variables the Hamiltonian reads
\bes
H = \sum_{n\in \Z} \left( {1\over 2} E_{-n}^aE^{n}_a +{i n\over2}
u_{-n-p}^i u^{n+p}_i +\sum_{m\in Z}\rho^a_{ij}A_a^{-n-m}u^i_{n-p}u^j_{m+p}
\right)\;\; ,
%\left\{ \begin{array}{cc}p= 0& R-case\\p={1\over 2}& NS-case \end{array}
%\right. ,
\ees
The algebra of the constraints in this representation is given by
\bes
\{G_a^n,G_b^m\} = f_{ab}^cG_c^{n+m}\label{km0}
\ees
and we also have
\bes
\{G_a^n, H\} = 0\;\;.
\ees
Eq.~(\ref{km0}) shows that the constraints generate a Kac-Moody algebra with
vanishing central charge which is a consequence of the fact that the
constraints are first-class, corresponding to the gauge symmetry of the system.

\section{Canonical Quantization of 2d YM-Theory}
In order to construct a quantum theory for this classical system one has to
replace the classical variables by operators and the fundamental Poisson
bracket relations become (anti-) commutators. Since at the quantum level the
fields are represented by
operators, one has to specify the space on which they act. This can be
achieved by a choice of vacuum because all other states can then be created
from
the vacuum. For the fermions there is not much freedom for the choice of
vacuum because the Hamilton operator has to be bounded from below.
A suitable vacuum can be defined by its property
\bes
u_n^i\vacr = 0 \;\; ,\;\; n > 0\;\; .
\ees
In the NS-case we can take the vacuum to be a non-degenerate single state,
whereas in the R-case the vacuum has to carry a representation of the
Clifford-algebra, generated by the zero modes of the fermions. As a
consequence of the condition (\ref{clherm}) on the classical variables,
which has to be respected by the inner product of the quantum theory, we
have
\bes
\vacl u_{-n}^i = 0 \;\; ,\;\; n > 0\;.
\ees
The physical picture of this is that the ket-states contain only right
moving particles and the bra states only left movers.

For the YM sector we first note that the operators $A_n^a,E_m^b$ do not
have the correct relations under Hermitian conjugation for the interpretation
as
creation and annihilation operators.
We therefore have to transform them to a new set of canonical operators.
A convenient choice is
\bes
\Phi^a_n = {1\over \sqrt{2}}(A_n^a + iE_n^a) \;\;, \;\;
\overline{\Phi}^a_n = {1\over \sqrt{2}}(A_n^a - iE_n^a)\;\;,
\ees
which leads to the following commutation relation for the new operators
\bes
[\Phi_n^a, \Phi^b_m] = 0 \;\; ,\;\;
[\Phi_n^a, \overline{\Phi}^b_m] = \delta^{ab}\delta_{n,-m}\;\; ,\;\;
[\overline{\Phi}_n^a, \overline{\Phi}^b_m] = 0\;\;. \label{iprym1}
\ees
With this definition we see that the reality condition (\ref{clherm}),
in terms of the new variables is
\bes
{\Phi^a_n}^\dagger =\overline{\Phi}^a_{-n}\; .
\ees
The Hamiltonian for the YM-fields becomes in this variables
\bes
H_{YM}= { 1\over 2} \sum_{n \in \Z} E^a_{-n}E^{n}_a = {1\over 4} \sum_{n\in \Z}
\left(\Phi^a_{-n}\Phi^n_a + \Phi^a_{-n}\overline{\Phi}^n_a +
\overline{\Phi}^a_{-n}\Phi^n_a + \overline{\Phi}^a_{-n}\overline{\Phi}^n_a
\right)
\ees
{}From this we see that again the boundedness of the Hamiltonian restricts the
possible choices for the vacuum. However, a vacuum which separates the right
and left moving fields again allows for a bounded Hamiltonian. We therefore
define the vacuum for the YM-fields
\bes
\begin{array}{cccccc}
\Phi^a_n\vacr &= 0 &,\; n \geq 0\;\; ,&\;\; \overline{\Phi}^a_n\vacr & = 0&,\;
 n\geq 1\;,\\ & & & & & \\
\vacl\Phi^a_{-n} &= 0&, \; n \geq 1\;\; ,&\;\; \vacl\overline{\Phi}^a_{-n}&
 = 0&,\; n\geq 0 \;.
\end{array}\label{iprym2}
\ees

The choice of a vacuum also implies a normal ordering prescription for
the operators which says in our case that all positive frequencies have to
be moved to the right. It is this normal ordering prescription which changes
the algebra of constraints. Indeed, we find for the constraints, which read
in the new variables\footnote{$:\; :$ denotes the the normal ordered product
of operators}
\bes
G_a^n = {n\over\sqrt{2}}(\Phi^n_a -\overline{\Phi}^n_a)
+ \sum_{r\in \Z}if_{ab}^c :\Phi^b_{-r}\overline{\Phi}^{r+n}_c:
- \sum_{r\in \Z}{i\over 2} \rho_{aj}^i :u_i^{n-r-p}u_{r+p}^j:\;\; ,
\ees
that their commutator algebra differs from the classical Poisson bracket
algebra by a central charge:
\bes
[G_a^n,G_b^m] = i f_{ab}^c G^{n+m}_c +\alpha_T n \delta^{ab}\delta_{n,-m}\;\; .
\label{fkm}
\ees
The central charge
\bes
\alpha_T = -\aYM + {1\over 2}\alpha_M \label{alphat}
\ees
is given by
\bes
\aYM \delta_{ab} = -f_{ac}^d f_{bd}^c > 0 \;\; , \;\;
\alpha_M \delta_{ab} = -\rho_{a\: j}^i \rho_{b\: i}^j > 0\;\; .
\ees
Thus we see, that the contribution of the YM-fields to the anomaly has a
negative
sign and its absolute value is the index of the adjoint representation of
$\cal G$, whereas the fermions contribute with one half of the positive value
of the index of their representation of $\cal G$.
Note, that if $2 \aYM > \alpha_M$ the total central charge is negative,
i.e. in the case of pure YM-theory the constraints form a Kac-Moody algebra
with negative central charge.

This central charge destroys the first-class nature of
the constraints and all constraints with non-zero Fourier index have mutated
to second-class constraints. However, they have a natural complex structure,
induced by the reality condition (\ref{clherm}), which is valid also
at the quantum level. This complex structure relates the positive and
negative Fourier modes:
\bes
{G_a^n}^\dagger = G^{-n}_a \label{colo}
\ees
Therefore we can apply the Gupta-Bleuler method and impose the constraints
with positive Fourier index on the ket states
\bes
G^n_a\ket = 0 \;\; n\geq 0 \;\; . \label{gbcon}
\ees
Hermitian conjugation leads to
\bes
\bra G_a^{-n} = 0 \;\;  n\geq 0 \;\; ,
\ees
and therefore eq. (\ref{gbcon}) guarantees the vanishing of all expectation
values of the constraints on physical states, i.e. we have for all physical
states
\bes
\bra G_a^n \ket = 0 \;\; n\in \Z\;\;.
\ees
Note, that all physical states are highest weight vectors of the Kac-Moody
algebra which are singlets under the action of the finite dimensional
algebra. The fact that Kac-Moody algebras with negative central charge have
no unitary highest weight representations therefore is not a problem in
principle since at most one state out of each representation is a physical
state.

\section{BRST-Quantization of Holomorphic Constraints}
In the usual BRST-quantization procedure for first class constraints
one constructs a Hermitian
nilpotent BRST operator $\Om$. The physical states are than given as
the solutions of
\bes
\Om \ket = 0\;,
\ees
i.e. physical states are BRST-closed.
However, the physical states are only defined up to some equivalence, i.e.
two states are equivalent when they differ by a BRST-exact term:
\bes
\ket\: \sim\, \ketp \;\eqv \; \ket  = \ketp +\: \Om \ketl
\ees
Using that $\Om$ is a Hermitian operator one can easily check that BRST-exact
states have zero norm and that the inner product does not depend on the choice
of a representative out of a equivalence class

In order to construct the BRST operator one introduces for each constraint
a pair of anti-commuting
Hermitian variables $c^a, \;b_a$ with the anti-commutation relation
\linebreak $[c^a,b_b]_+=\delta^a_b$.
Quantities depending on the ghosts can be labeled by their
ghost number which is defined on the ghost variables as
\bes
\gh{c^a} = 1\;\;,\;\; \gh{b_a} = -1\;.
\ees
The BRST-operator $\Om$ then takes the form
\bes
\Om = c^aG_a -{1\over2} c^a c^b f_{ab}^c b_c + ...\;\;\;\mbox{with}\label{brst}
 \\
\Om^2 = 0\;\;\;\mbox{and}\;\;\; \gh{\Om} = 1\;,
\ees
where the $G_a$ are the Hermitian first-class constraints and the higher order
ghost terms occur when the algebra is open and/or the constraints are reducible
(for a detailed review of this construction see \cite{Henn}). The BRST-closed
states with ghost number zero fulfill the Dirac condition
\be
G_a\ket = 0\; \; .
\ee

Let us now turn to a system which has in addition to the Hermitian first-class
constraints
$G_a$ some second-class constraints which allow for a holomorphic split at the
quantum level, i.e. the second-class constraints are given by $G_r$ and
$G_r^\dagger$ and their commutation relations read
\be \begin{array}{rcl}
[G_a , G_b ] &=& f_{ab}^cG_c \\
 & & \\
{}[ G_r , G_a ] &=& f_{ra}^tG_t + f_{ra}^bG_b\\
 & & \\
{}[ G_r , G_s ] &=& f_{rs}^tG_t + f_{rs}^aG_a\\
 & & \\
{}[ G_r , G_s^\dagger ] &=& Z_{rs} + h_{rs}^tG_t + h_{rs}^{\prime t}G_t^\dagger
\;. \end{array}
\ee
Since the
constraints are not Hermitian, a BRST-operator constructed as $\Om$ in
eq (\ref{brst}) cannot be Hermitian and therefore the BRST-exact states
do not necessarily have zero norm:
\bes
|\Lambda^\Om\!\!> = \Om\ketl \; ,\;\;
\| |\Lambda^\Om\!\!>\!\!\!\|^2 = \:  \bral\Om^\dagger\Om\ketl
\ees
This can be understood in the following way. In the case of first-class
constraints the constraints generate symmetry transformations on the constraint
surface. This symmetry is also present in the quantum theory and leads to
equivalence classes in the BRST-cohomology, where different elements of one
equivalence class differ by a zero norm state, i.e. by a symmetry
transformation.
In the case of second-class constraints there are no such symmetry
transformations. Although one can split the constraints into two sets such
that the commutator on each set is in involution, one cannot interpret
the elements of one set as symmetry transformations, since the implementation
of one set on the the ket-states automatically implements the other set on
the bra-states via Hermitian conjugation. In other words, an implementation
of the symmetry on the ket-states introduces also a gauge fixing on the
bra-states.

However, a scenario where a non-Hermitian BRST-operator can be
used to introduce a gauge fixing of first-class constraints has been disussed
in \cite{JW,JWW}. This is achieved by changing the inner product for the ghost
variables such that $Im \Om$, i.e. the space of BRST-exact states, contains no
degenerate states in this new inner product space. With this inner product
$\Om$ and the co-BRST-operator $\Om^\dagger$ are different and the gauge
fixed states are found
as the zero modes of the non trivial BRST-Laplacian $\Delta = \Om^\dagger
\Om + \Om\Om^\dagger$, or equivalently as BRST-closed and co-BRST-closed
states.
It has been shown that a non-degenerate inner product
on the subspace $Im \Om$ or equivalently $Ker \Om$ is equivalent with a
complete gauge fixing.

We have already seen that the BRST-operator for holomorphic constraints is
not Hermitian and that we therefore cannot expect the BRST-exact
states to be of zero norm.
The holomorphic nature of the constraints together
with a natural choice for the inner product guarantees that we have
a nilpotent BRST-operator $\Om$ together with a nilpotent co-BRST-operator,
such that the BRST-Laplacian $\Delta$ is non-trivial.

In order to make this more precise we have to introduce the ghosts
for the holomorphic and anti-holomorphic constraints. Since the constraints
are non-Hermitian it appears natural to choose the ghosts such that they have
the same complex structure as the constraints. Hence we define k
pairs of ghosts $\eta^r, \etb_r$ such that $\eta^r$ and $\etb_r$ are
related by Hermitian conjugation\footnote{This is different form the choice
introduced by Kugo and Ojima \cite{KO} since we are dealing with second-class
constraints and gauge fixed first-class constraints.}.
The ghost numbers are assigned as usual, i.e.
$\gh{\eta^r}=1$ and $\gh{\etb_r}=-1$. An inner product which relates the ghosts
and ghost momenta via Hermitian conjugation can be defined by using the Hodge
duality on the ghosts (details of this definition can be found in \cite{JW}).
The anti-commutation relations are defined
as
\bes
[\eta^r, \eta^s]_+ = 0\;\; ,\;\; [\etb_r,\etb_s]_+ = 0
\;\;,\;\;[\eta^r,\etb_s]_+ = \delta^r_s\;.
\ees
Since Hermitian conjugation changes the ghost number of the holomorphic ghosts,
the ghosts $c^a , b_b$ for the first-class constraints also have to be related
by Hermitian conjugation. Otherwise the co-BRST-operator is not an eigenstate
of
the ghost number operator.
With this ghost system the BRST-operator for the holomorphic constraints
is constructed as usual and (to cubic order in the ghosts) takes the form
\bes
\Om &=& c^aG_a + \eta^rG_r -  \\
&-& {1\over 2} c^ac^bf_{ab}^cb_c -
{1\over 2} \eta^rc^af_{ra}^bb_b - {1\over 2} \eta^rc^af_{ra}^s\etb_s -
{1\over 2} \eta^r\eta^sf_{rs}^t\etb_t - {1\over 2} \eta^r\eta^sf_{rs}^ab_a +
...
\nonumber
\ees
The co-BRST-operator is obtained by Hermitian conjugation,
\bes
\Om^\dagger &=& b^aG_a + \etb^rG^\dagger_r -  \\
&-& {1\over 2} b^ab^bf_{ab}^cc_c -
{1\over 2} \etb^rb^af_{ra}^bc_b - {1\over 2} \etb^rb^af_{ra}^s\eta_s -
{1\over 2} \etb^r\etb^sf_{rs}^t\eta_t - {1\over 2} \etb^r\etb^sf_{rs}^ac_a +
...
\nonumber
\ees
The BRST-Laplacian is then
\bes
\Delta = [\Om^\dagger,\Om]_+ = G_r^\dagger G^r + G_aG^a + \mbox{\it ghost
terms}
\ees
and the ghost numbers of these operators are
\bes
\gh{\Om} = 1\;\;,\;\;\gh{\Om^\dagger}=-1\;\;,\;\;\gh{\Delta}=0\;.
\ees
The ghost vacuum is chosen such that it is annihilated by the ghost momenta.
The physical states are given by the zero modes of the BRST-Laplacian. The
zero modes are also BRST-closed and co-BRST-closed, i.e. we have for
physical states:
\bes
\Om\ket = 0 \;\; \mbox{and}\;\; \Om^\dagger\ket = 0\;. \label{gf}
\ees
BRST-closed states with ghost number zero are  solutions of
\bes
G_r|\Psi^{(0)}\!>\; = 0\;\; \mbox{and}\;\; G_a|\Psi^{(0)}\!> = 0\;.
\ees
On the other hand all states with ghost number zero are co-BRST-closed
since $\Om^\dagger$ has $\gh{\Om^\dagger}=-1$. This shows that there is a
Gupta-Bleuler realization of the constraints on the zero modes of the
BRST-Laplacian with ghost number zero. For further details see \cite{wk}.

The choice of the ghost system for the holomorphic constraints and the
requirement that the co-BRST operator is an eigenstate of the ghost number
operator has led to non-Hermitian ghost system for the first-class constraints.
This corresponds to a gauge-fixing of the first class-constraints
\cite{JW,JWW}.
This point will be discussed in some more detail in the case of
2d-Yang-Mills Theory in section 6.

Before we construct the BRST operator for 2d YM-theory we
want to make a remark about the possible limitations of the general
construction. Suppose an anomalous system admits a complex structure such
that the constraints can be split into holomorphic and anti-holomorphic
constraints and we have performed the BRST construction described in this
section. Although we have seen that for the cohomology class at ghost level
zero there is a Gupta-Bleuler type realization of the constraints on the
states,
it is not
guaranteed that this defines a consistent physical quantum theory. E.g. it
is not clear that the inner product on the cohomology class at ghost level
zero or any other ghost level is positive definite. Also the renormalizability
of such a theory in more than 1 space-dimension is an open problem. However,
it has been shown that anomalous YM-theories coupled to chiral fermions in
1 space-dimension can be quantized via the Gupta-Bleuler method, see
\cite{JMi,MbP,HeHo} and references therein. Therefore we may expect that
for such a system a BRST quantization along the lines described in this section
will lead to meaningful physical results.

\section{The Ghost System for 2d YM Theory}
{}From the algebra of the constraints (\ref{fkm}) for 2d YM theory we see that
the zero modes
of the constraints remain first-class at the quantum level. They generate the
finite algebra $\cal G$. Furthermore, the constraints with positive
Fourier index generate a closed subalgebra which is first-class. They represent
the holomorphic constraints in our example since they are related to the
negative modes of the constraints by Hermitian conjugation. The BRST operator
corresponding to a Gupta-Bleuler quantization of this system should therefore
only contain the constraints $G_a^n$ with $n \geq 0$. As in the previous
section,
we introduce pairs of ghosts $c_{-n}^a,\cb_m^b$ but this time only for
$n\geq 0$, i.e. for the zero modes and positive modes of the constraints.
Again the ghosts have the standard anti-commutation relations
\bes
[c^a_{-n},c^b_{-m}] = 0\;\; ,\;\;[c^a_{-n},\cb^b_{m}] = \delta^{ab}\delta_{n,m}
\;\; ,\;\; [\cb^a_{n},\cb^b_{m}] = 0\;\;.
\ees
We already saw in section 4 that a natural choice for the complex structure of
the ghosts for the holomorphic constraints is that they obey the same
relations under Hermitian conjugation as the holomorphic and anti-holomorphic
constraints. We therefore have
\bes
{c_{-n}^a}^\dagger = \cb_n^a \; ,\; n> 0 \;\;.
\ees
It is now only natural to extend this complex structure to the ghosts for
the first-class constraints
\bes
{c_{0}^a}^\dagger = \cb_0^a  \;\;.
\ees
The canonical choice of the ghost vacuum is
\bes
\cb^a_n\vacr = 0 \; ,\; n\geq 0\;\;.
\ees
With this definition the inner product on the ghost space is positiv definite.

We can define a representation of the algebra of constraints on the ghost
states
\bes
T_a^n = -i \sum_{r\geq 0} f_{ab}^c c_{-r}^b\cb^{r+n}_c\;\; n\geq 0\;\; ,\\
T_a^{-n} = -i \sum_{r\geq 0} f_{ab}^c c_{-r-n}^b\cb^{r}_c\;\; n> 0\;\; .
\ees
However, these operators do not form a Kac-Moody algebra since these ghosts
are constructed only for the zero modes and positive modes of the constraints.
Therefore we have
\bes
[T^n_a, T^m_b ] = i f_{ab}^c T_c^{n+m}\;\;n,m\geq 0\;\; ,
\ees
\bes
[T^{-n}_{a}, T^{-m}_{b} ] = i f_{ab}^c T_c^{-n-m}\;\;n,m\geq 0\;\; ,
\ees
but
\bes
[T^{n}_a,T^{-m}_b]= i f_{ab}^cT_c^{n-m} -
\left\{ \begin{array}{cc}
{1\over 2}
\sum_{r=0}^{m-1}f_{ab}^c f_{cd}^e c^{-s}_e \cb_{s+n-m}^d & n\geq m \\ & \\
{1\over 2}
\sum_{r=0}^{n-1}f_{ab}^c f_{cd}^e c^{-s-m+n}_e \cb_{s}^d & n< m\end{array}
\right. \;\;n,m\geq 0\;\; . \label{fail}
\ees
We see that the operators $T^n_a$ generate the algebra of the holomorphic
resp.~anti-holomorphic and the first-class constraints, but they fail to
generate the full Kac-Moody algebra. However, those operators obey the same
relations under Hermitian conjugation as the constraints
\bes
{T^n_a}^\dagger=T^{-n}_a\;\; .
\ees

With the definitions given above we can now construct the BRST
operator. The BRST operator corresponding to a Gupta-Bleuler quantization
is given by
\bes
\Om =\sum_{n\geq 0} c_{-n}^a(G_a^n +{1\over 2}T^n_a)\;\; .\label{omf}
\ees
One can check that this operator is nilpotent, i.e.
\bes
\Om^2=0\;\; .
\ees
According to the complex structure of the ghosts and the constraints the
co-BRST operator, the adjoint operator of $\Om$, is given by
\bes
\Omb =\sum_{n\geq 0} (G_a^{-n} +{1\over 2}T^{-n}_a)\cb_n^a\;\; .
\ees
Like the BRST operator, this operator is also nilpotent
\bes
{\Omb}^2=0\;\; .
\ees
We note that the nilpotency of both the BRST $\Om$ and the co-BRST operator
$\Omb$ does not depend on the central charge of the Kac-Moody algebra.

\section{The BRST-Laplacian}
We can now use the nilpotent BRST operator $\Om$ and its adjoint $\Omb$
to construct the BRST-Laplacian
\bes
\Delta = [\Omb,\Om] =
{1 \over2}(G^0_a + T^0_a)^2 + {1\over 2}(2\alpha_T +\aYM)(L_0 + D_0)
\;\; ,  \label{GBlp}
\ees
where
\bes
D_0 = \sum_{n > 0} n c_{-n}^a \cb_a^n
\ees
is the ghost momentum operator and
\bes
L_0 = {1\over 2\alpha_T +\aYM } (G^a_0 G^0_a + 2 \sum_{n\geq 1} G^a_{-n}G^n_a)
\label{Lnull}
\ees
stands for the energy momentum operator of the Virasoro
algebra obtained via the Sugawara construction. The generators of this
algebra are given by
\bes
L_n = {1\over 2\alpha_T +\aYM }
\sum_{r\in\Z}\noSl G^a_{-r}G^{r+n}_a\noSr\;\; . \label{sugawara}
\ees
Here $\noSl\; \noSr$ denotes the normal ordering
\bes
\noSl G^n_aG^m_b\noSr = \left\{ \begin{array}{cc}
G^n_aG^m_b & n\leq m\\ & \\
G^m_bG^n_a & n > m \end{array}\right.\;\;\; .
\ees
The commutation relations of the Virasoro generators with the constraints
are
\bes
[L_n,G^m_a] = -mG^{m+n}_a\label{KacVir}
\ees
and the commutation relations of the Virasoro generators with themselves are
\bes
[L_n,L_m]=(n-m)L_{n+m} + {\alpha_T dim{\cal G} \over 6(2\alpha_T +\aYM)}
n(n^2-1)\delta_{n,-m}\;\; .
\ees
It is straightforward to compute the commutator of the Virasoro generators with
the BRST-Laplacian and we obtain
\bes
[\Delta, L_n] = -{n\over 2}(2\alpha_T + \aYM ) L_n\;\;.\label{suvira}
\ees
We can now also try to construct a Virasoro algebra for the ghosts. However,
the definition
\bes
D_n = \left\{ \begin{array}{cc}
\sum_{r\geq 0} r c^a_{-r}\cb^{r+n}_a & n\geq 0\\ & \\
\sum_{r\geq 0} r c^a_{-r+n}\cb^{r}_a & n< 0\end{array}\right. \;\;\; ,
\ees
does not lead to the desired result, since only the commutator of generators
which have positive resp.~negative indices gives the correct result
\bes
[D_n,D_m]=(n-m)D_{n+m} \;\;\; n m\geq 0
\ees
whereas the commutator $(n,m\geq 0)$
\bes
[D_n,D_{-m}]=(n+m)D_{n-m} - \left\{ \begin{array}{cc}
m^2 \sum_{r\geq 0} c_{-r}^a\cb_a^{r+m-n} -
\sum_{r=0}^{m-1}c_{-r}^a\cb_a^{r+n-m}
& n\geq m\\ & \\
n^2 \sum_{r\geq 0} c_{-r+n-m}^a\cb_a^r - \sum_{r=0}^{n-1}c_{-r+n-m}^a\cb_a^r
& n < m \end{array}\right. \;\;
\ees
fails to give the commutation relation of a Virasoro algebra. This result is
similar to the one we obtained in eq. (\ref{fail}). However we find for the
commutator of the $D_n$ with the BRST-Laplacian
\bes
[\Delta, D_n] = -{n\over 2}(2\alpha_T + \aYM ) D_n\;\;\; n\in \Z \;\;.
\ees

We note that all terms of the BRST-Laplacian given by (\ref{GBlp}) commute
with each other and we can therefore diagonalize them simultaneously. The
spectrum of the BRST-Laplacian is then given by the combined spectrum of its
summands. The first term of (\ref{GBlp}) is the quadratic Casimir operator
of the finite dimensional algebra $\cal G$ in a representation acting on
the extended state space (i.e. it also acts on the ghosts). The spectrum
of this operator is positive. The last term is the momentum operator and
the sign of its spectrum is determined by the sign of the prefactor
$2\alpha_T+\aYM$ which is negative for a pure YM-system. In this case the
spectrum is negative. Also the spectrum of the second term $\sim L_0$
depends on the sign of $2\alpha_T +\aYM$. It is positive and bounded by zero
for $2\alpha_T+\aYM >0$ and it is unbounded for $2\alpha_T + \aYM < 0$.
Therefore we expect that the BRST-Laplacian (\ref{GBlp}) does not necessarily
lead to a gauge fixing of the first class-constraints $G_a^0$ in the case
$2\alpha_T +\aYM < 0$. A positive contribution of the Casimir acting on a
non-trivial representation of $\cal G$ can be cancelled by a negative
contribution of $(2\alpha_T +\aYM)(L_0 + D_0)$. In fact, as we will show below,
the BRST-operator $\Om$ can be written as a sum of two commuting nilpotent
operators $\Om_+$ and $\Om_0$ where the spectrum of the Laplacian for $\Om_0$
is always positive definite and the Laplacian of $\Om_+$ contains the
indefinite
part of $\Delta$. From this we conclude that the indefinite part of $\Delta$
and the Casimir operator have to vanish seperately for gauge fixed states.
Otherwise they would be in a doublet representation of the algebra generated
by $\Om_+ , \Om_+^{\dagger}$ resp.~$\Om_0 , \Om_0^{\dagger}$ and therefore
they would also be in a doublet representation of the algebra generated by
$\Om ,\Om^{\dagger}$ \cite{JWW}.

In order to remove the gauge symmetry from $Ker \Delta$ let us study in a
little more detail the relation of the first-class resp.~second-class
constraints to the BRST-Laplacian. The BRST-operator (\ref{omf}) has been
constructed for both the first-class and second-class constraints. However,
we also can consider a nilpotent operator containing only second-class
constraints,
\bes
\Om_+ =\sum_{n \geq 1} c_{-n}^a ( G_a^n -{i\over 2} \sum_{m\geq 1}
f_{ab}^c c_{-m}^b\cb_c^{n+m} ) \;\;\; .
\ees
One can check that this operator is nilpotent
\bes
\Om^2_+=0\;\; .
\ees
This operator defines a cohomology only on the second-class constraints and
does
not take into account the first-class constraints $G_a^0$. For those we can
construct the nilpotent operator
\bes
\Om_0 &=& c^a_0 ( G^0_a -{i\over 2} f_{ab}^c c^b_0\cb_c^0 + \tilde{G}_a )
\label{om0}
\ees
with
\bes
\tilde{G}_a= -i\sum_{n\geq 1}f_{ab}^c c^b_{-n}\cb^n_c \;\;\; , \;\;\;
\tilde{G}_a^\dagger =\tilde{G}_a
\ees
and
\bes
\Om_0^2&=& 0\;\;.
\ees
This operator differs from the usual BRST operator for a gauge algebra $\cal G$
generated by the $G_a^0$'s by the last term $\tilde{G}_a$ in the bracket in eq.
(\ref{om0}) which is due to the fact that the first-class and second-class
constraints do not commute. This term defines a representation of $\cal G$ on
the ghosts for the second-class constraints. It also guarantees that
\bes
[\Om_+ ,\Om_0 ] = 0\;\; .
\ees
The adjoints of $\Om_+$ and $\Om_0$ are given are given by
\bes
\Om_+^\dagger &=&\sum_{n \geq 1} ( G_a^{-n} -{i\over 2} \sum_{m\geq 1}
f_{abc} c_{-m-n}^b\cb^c_m ) \cb^c_n \;\; ,\\
\Om_0^\dagger &=& ( G^0_a-{i\over 2} f_{abc}c^b_0\cb^c_0 + \tilde{G}_a)\cb^a_0
\;\; .
\ees
With those operators we now can construct the BRST-Laplacian for the
first-class constraints,
\bes
\Delta_0 := [\Om_0^\dagger ,\Om_0 ] = {1\over 2} (G^0_a + \tilde{G}_a)^2 +
{1\over 2} (G^0_a + \tilde{G}_a - {i\over 2} f_{ab}^c c^b_0\cb_c^0 )^2\;\; .
\ees
This operator is the sum of two squares of Hermitian operators and therefore
we have
\bes
\Delta_0\ket = 0\;\;\; \LRA\;\;\;
( G^0_a +\tilde{G}_a ) \ket = 0 \;\; \wedge\;\; -{i\over 2}f_{ab}^cc^b_0\cb_c^0
\ket = 0 \;\;.\label{finsol}
\ees
The operator $\Delta_0$ is equivalent to the BRST-Laplacian studied in
\cite{JW}. There it was shown that it completely fixes the symmetry generated
by
the algebra $\cal G$, i.e. $Ker \Delta_0$ contains only one state out of each
set of gauge equivalent states.

The BRST-Laplacian which is related to $\Om_+$ is given by
\bes
\Delta_+ := [\Om_+^\dagger , \Om_+ ] =
- \tilde{G}^aG_a^0 - {1\over 2} \tilde{G}^a\tilde{G}_a + \sum_{n\geq 1}
( G^a_{-n}G^n_a + {1\over 2} ( 2\alpha_T + \aYM ) n c^a_{-n}\cb^n_a ) \;\; .
\ees
Computing the anti-commutator of $\Om_0$ with $\Om_+^\dagger $ we find
\bes
[ \Om_0 , \Om_+^\dagger ] = 0
\ees
and therefore we can write the full BRST-Laplacian as
\bes
\Delta = \Delta_0 + \Delta_+
\ees
with
\bes
[ \Delta_0 , \Delta_+ ] = 0 \;\; .
\ees
{}From this we conclude that the representatives of the cohomology classes of
the full BRST-operator $\Om$ are those states which are simultaneous solution
to
\bes
\Delta_0 \ket = 0 \label{cosol1}
\ees
and
\bes
\Delta_+ \ket = 0 \;\; . \label{cosol2}
\ees
Furthermore we can rewrite $\Delta_+$ restricted to $Ker \Delta_0$ as
\bes
\Delta_+|_{\Delta_0 = 0} = {1 \over 2} (2\alpha_T + \aYM) ( L_0 + D_0 ) \;\;.
\label{suvir}
\ees
We see  that $\Delta_+|_{\Delta_0=0}$ takes the simple form of the energy
momentum
operator and hence $\Om_+$ and $\Om_+^\dagger $ are in some sense the
``square-roots'' of this operator.

Moreover we have seen that although the full BRST-Laplacian does not fix a
gauge for the first-class constraints one can split this operator into two
parts
corresponding to the first-class and second-class constraints such that the
intersection of the kernels of the two new operators contains only gauge
fixed states. A similar method has been used in \cite{BPPFM} for string theory.

\section{The Cohomology of $\Om$}
The cohomology classes of $\Om$ are represented by solutions to the
eqs.~(\ref{cosol1}) and (\ref{cosol2}), i.e. we have split the full BRST
symmetry generated by $\Om$ into
two parts such that we can now study the cohomology problem of two different
BRST operators $\Om_0$ and $\Om_+$. Fixing a gauge for one of them (we
will choose $\Om_0$) we will still have the BRST-symmetry generated by the
other
operator ($\Om_+$). In other words, there is a BRST-symmetry generated by
$\Om_+$ on the space of representatives for the cohomology-classes of $\Om_0$.
One has to choose a second gauge fixing, this time for $\Om_+$, in order to
remove the remaining BRST-symmetry.

A complete gauge fixing for $\Om_0$ is
given by eq.~(\ref{cosol1}),
which was studied in ref.~\cite{JW} where it was found that the solutions are
states which are singlets of the algebra of first-class constraints, which in
our case is generated by $G^0_a + \tilde{G}_a$, i.e. the representatives of the
cohomology of $\Om_0$ are given as solutions to (\ref{finsol}). For a
characterization of these solutions it is convenient to distinguish between the
ghosts for the first-class constraints, which have zero momentum, i.e. the
pairs $(c^a_0 ,\cb_a^0 )$ and the ghost for the second-class constraints with
non-zero momentum, the pairs $( c^a_{-n}, \cb^n_a)$, $n\geq 1$. Let $\Hz{gh_0}$
denote the space spanned by the $c^a_0$. The full extended state space $\Hext$
can than be written as
\bes
\Hext = \Hz{1}\otimes\Hz{gh_0}\;\; .
\ees
We now see that the two eq.~(\ref{finsol}) are conditions on states in
different
spaces, i.e. we have
\bes
( G^0_a +\tilde{G}_a ) \ket = 0\;\;\; ,\;\;\; \ket\in\Hz{1}\label{nacht1}
\ees
and
\bes
-{i\over 2}f_{ab}^cc^b_0\cb_c^0 \ketp = 0 \;\;\;, \;\;\;\ketp\in\Hz{gh_0}\;\;.
\label{nacht2}
\ees
Solutions of (\ref{nacht2}) are of course the vacuum state and the state
with all ghost levels filled. Further solutions are given by states where
the algebra indices of the ghosts are contracted by total
antisymmetric tensors which are invariant under the action of $\cal G$, e.g.
the state
\bes
\ketp = f_{abc}c^a_0c^b_0c^c_0\vacr
\ees
is such a solution. Let us denote the space of solutions to (\ref{nacht2})
by $\Hz{gh_0}^\prime$.

The space $\Hz{1}$ can be written as a tensor product of the original state
space $\Hz{0}$, i.e. the state space without ghosts, and the space spanned by
the ghosts with non-zero momentum $\Hz{gh_+}$:
\bes
\Hz{1} = \Hz{0}\otimes\Hz{gh_+}\;\; .
\ees
However, eq.~(\ref{nacht1}) cannot be split any further and the statement
that the solutions of (\ref{nacht1}) are singlets under the action of $\cal G$
does not imply that the states in $\Hz{0}$ are singlets unless all ghost levels
in $\Hz{gh_+}$ are empty. Of course, there are also non-trivial ghost-states on
which $\cal G$ acts trivially, e.g. the invariant tensors which lead to
solutions of (\ref{nacht2}) can be used to construct such states in
$\Hz{gh_+}$.
Since ghosts can have different momenta there are even more possibilities
to construct singlet states in $\Hz{gh_+}$ e.g.
\bes
\ket = \delta_{ab}c^a_{-n}c^b_{-m}\vacr
\ees
is such a state. However, in general the ghost-states do not transform
trivially
under $\cal G$ but they are in some representation of $\cal G$ which can be
generated by taking anti-symmetric {\it and} symmetric tensor products of the
adjoint representation of $\cal G$.

The condition on a state
\bes
\ket = \ketl\ketph\;\;\; , \;\;\;\ketl\in\Hz{0}\; ,\;\ketph\in\Hz{gh_+}
\ees
to be a solution to (\ref{nacht1}) is that both $\ketl$ and $\ketph$ are in
the same representation of $\cal G$ such that they are together invariant
under the action of $\cal G$. In the case of pure YM-theory, where the states
of $\Hz{0}$ are generated by operators which are in the adjoint representation
of $\cal G$, any state in $\Hz{0}$ is in a representation which can be
generated
from tensor products of the adjoint representation and hence there is a ghost
state in $\Hz{gh_+}$ such that they together form a singlet.
Conversely, we can infer from the fact that the state space of pure YM-theory
is always contained in $\Hz{0}$ that all states in $\Hz{gh_+}$ appear as
solutions to (\ref{nacht1}). The situation is
different in the case of YM-theory coupled to matter. If the matter fields
are in the fundamental representation of $\cal G$ then the states of $\Hz{0}$
appear in all representations of $\cal G$. But not all representations can be
obtained by taking tensor products of the adjoint representation. Therefore in
this case not all states of $\Hz{0}$ appear as solutions to (\ref{nacht1}).
Let us denote by $\Hz{1}^\prime$ the space of
solutions in $\Hz{1}$.

So far we have considered only the cohomology of $\Om_0$ and we have given
a description of the space of representatives of its cohomology classes. In
order to find representatives of the cohomology classes of the full BRST
operator $\Om$ we have to fix a gauge for the remaining BRST symmetry generated
by $\Om_+$. This can be done by imposing the vanishing of $\Omb_+$ on the
states. We first note that $\Delta_+$ acts trivially on $\Hz{gh_0}^\prime$:
\bes
\Delta_+\ket =0 \;\; ,\;\; \forall \ket \in \Hz{gh_0}^\prime\;\; ,
\ees
from which we infer that the general structure of the space of solutions
$\Hz{sol}$ to (\ref{cosol1}) and (\ref{cosol2}) is given as
\bes
\Hz{sol} = \Hz{sol2}\otimes\Hz{gh_0}^\prime
\ees
with
\bes
\Hz{sol2} = \{ \ket\in\Hz{1}^\prime |\; \Delta_+\ket =0\}\;\; ;
\ees
i.e. we have
to look for solutions of (\ref{cosol2}) in the space of solutions of
(\ref{cosol1}). On this space eq.~(\ref{cosol2}) reduces to (\ref{suvir})
and it is simply the sum of the Virasoro energy-momentum operator and the
momentum operator of the ghosts. Depending on the value of $2\alpha_T +\aYM$,
which according to eq.~(\ref{alphat}) is
\bes
2\alpha_T +\aYM = -\aYM + \alpha_M
\ees
we can distinguish three different situations:
\begin{itemize}
\item[1.] $\alpha_M > \aYM $, which is generically the case if the
representation of the fermions coupled to the YM-fields is higher dimensional
than the adjoint representation of $\cal G$. In this case
the spectra of both, $(\alpha_M-\aYM )L_0$ and $(\alpha_M-\aYM )D_0$, are
positive and bounded
from below by zero. As a consequence both operators have to be zero separately.
$D_0$ acts only non-trivially on $\Hz{gh_+}$ and the only state on which $D_0$
vanishes is the trivial state. Therefore the space of solutions is of the
form
\bes
\Hz{sol} = \Hz{sol2}\otimes\Hz{gh_0}^\prime
\ees
with
\bes
\Hz{sol2} = \{ \ket\in\Hz{0} |\; \Delta_+\ket =0\ \; \wedge G^0_a\ket =0\}\;\;
{}.
\ees
We see that the only non-trivial ghost-states are contained in
$\Hz{gh_0}^\prime$
and that space of solutions in $\Hz{0}$ at non-zero ghost level is simply a
copy of the space of solution at ghost level zero. Thus the cohomology
generated
by $\Om$ is completely equivalent to the cohomology of $\cal G$ generated by
$\Om_0$.
\item[2.] $\alpha_M < \aYM$, which is generically the case if the
representation
of the fermions is lower dimensional than the adjoint representation. Now the
spectrum of  \linebreak
$(\alpha_M -\aYM)D_0$ is negative and bounded from above by zero. But
the spectrum of $(\alpha_M - \aYM )L_0$ is unbounded. In addition to the
solutions we found for the first case there are now also solutions where a
positive contribution of $(\alpha_M - \aYM )L_0$ acting on a state in $\Hz{0}$
is cancelled by a negative contribution of $(\alpha_M -\aYM )D_0$ acting on a
state in $\Hz{gh_+}$, i.e. we have for solutions to (\ref{cosol1}) and
(\ref{cosol2}) in $\Hz{1}$
\bes
(G^0_a + \tilde{G}_a)\ket\!\ketph = 0\;\; \wedge\;\;
L_0\ket\!\ketph = -D_0\ket\!\ketph
\ees
with
\be
\ket\in\Hz{0}\; ,\;\ketph\in\Hz{gh_+}\;.
\ee
We will discuss this case in more detail for pure $SU(2)$-YM-theory below.
\item[3.] $\alpha_M =\aYM $, which is generically the case, if the fermions are
in the adjoint representation. In this case the ghost momentum operator $D_0$
vanishes in $\Delta_+$ because its prefactor becomes zero. Also the operator
$L_0$ cannot be defined as before since the prefactor in (\ref{Lnull}) is zero.
One has to define $L_0$ without that factor and $\Delta_+$ becomes than
\bes
\Delta_+ = {1\over 2}G^0_aG^a_0 + \sum_{n\geq 1}G_{-n}^aG_n^a \;\; .
\ees
We see that $\Delta_+$ imposes no further restrictions on $\Hz{gh_+}$ and hence
according to our discussion given above the entire space $\Hz{gh_+}$ appears
in the space of solutions to (\ref{cosol1}) and (\ref{cosol2}). This case is
also special from the point of view of representation theory of Kac-Moody
algebras since the Sugawara construction (\ref{sugawara}) also fails in this
case. One may define the Virasoro generators in (\ref{sugawara}) without the
prefactor, but the operators obtained in this way commute with the generators
of the Kac-Moody algebra. This case is called critical representation
\cite{mystery}.
\end{itemize}
So far we have discussed the general structure of the cohomology of $\Om$ for
a Kac-Moody algebra related to an arbitrary simple Lie group. We have found a
rich structure for negative values of the central charge
$\alpha_T < -{1\over2}\aYM$.

\section{The Cohomology of pure $SU(2)$-YM Theory}
We now wish to analyse in some more detail the cohomology of $SU(2)$-YM theory
without matter coupling. This is the only case for $SU(2)$-YM theory
which fits into case {\bf 2.} of the previous section (if one does not
allow for chiral bosons coupled to the YM-fields, which would contribute
with negative values to the central charge).

Since all fields of the theory (YM-fields and ghosts) are in the adjoint
representation, all representations of $SU(2)$ appearing in $\Hz{ext}$
can be labelled by a positive integer $k$, denoting the highest weight of that
representation.

As a first step we note that the only solutions in $\Hz{gh_0}$ are the states
with no ghosts or three ghosts. There is no non-trivial invariant
anti-symmetric tensor at ghost level 2.

We now turn to solutions of $\Delta_+$ in $\Hz{1}^\prime$, i.e. solutions in
the
space generated by the YM-fields and ghosts with non-zero momentum, which are
together singlets of $\cal G$. The representations of $\cal G$ on $\Hz{gh_+}$
and $\Hz{0}$ are products of the adjoint representation and therefore they are
the integer representations of $SU(2)$ labeled by a positive integer ${\bf k}$.
Let $\ket$ be a state in the representation ${\bf k}$, then the
quadratic casimir $C$ on this state is
\bes
C\ket = k(k+1)\ket\;\; .
\ees
{}From this we see also that with this normalization we have
\bes
\aYM = 2\;\; .
\ees
Let us denote by $\hws{k}\;$ a highest weight state of the $SU(2)$-Kac-Moody
algebra in $\Hz{0}$, which is in the representation ${\bf k}$ of the finite
dimensional $SU(2)$, i.e.
\bes
G_a^n\hws{k}\; = 0\; , n\geq 1\;\; ,
\ees
Since $\hws{k}$ is a highest
weight state the eigenvalue of $L_0$ for this state is given by the Casimir
operator and hence
\bes
\Delta_+\hws{k}\; = {k(k+1)\over 2}\hws{k}\;\; .
\ees
This eigenvalue of $\hws{k}\;$ has to be compensated by a corresponding
negative eigenvalue of a ghost state, which is also in the ${\bf k}$
representation of $SU(2)$. The ghost state with the smallest momentum for a
given (non-trivial) representation is the one with the smallest ghost number.
Hence the ghost state in representation ${\bf k}$ with smallest momentum is
given by a state with $k$ ghosts, where each of the first k momentum levels
is filled by one ghost. Let us denote such a state by $\hws{k,k}_{gh}\;$.
More generally, a state at ghost level $l$ in the representation ${\bf k }$
will be denoted by $\hws{l,k}\;$. The eigenvalue of $\Delta_+$ for this state
is given by
\bes
\Delta_+\hws{k,k}_{gh}\; = -{k(k+1)\over 2}\hws{k,k}_{gh} \;\; .
\ees
Thus we see that this state has the same eigenvalue with opposite sign as
$\hws{k}$.
Thus we infer that a highest weight state $\hws{k}$ in $\Hz{0}$ appears in the
cohomology of $\Om$ at ghost level k:
\bes
\Delta_+(\hws{k}\hws{k,k}_{gh})_0\; = 0 \;\; ,
\ees
where the subscript $0$ denotes $\cal G$ invariant components of the tensor
product.

We now turn to states in $\Hz{0}$, which are not highest weight states, but are
generated from them by applying $G^{-n}_a$ on a highest weight state. The
eigenvalues of $\Delta_+$ for such states coupled to ghost states such that
they are  singlets under $\cal G$ are (using (\ref{KacVir}) and
(\ref{suvir})) given by:
\bes
\Delta_+(G_a^{-n}\hws{k}\ket_{\!\!gh})_0 = ({k(k+1)\over 2} -n - n_{gh})
(G_a^{-n}\hws{k}\ket_{\!\! gh})_0\;\; , \label{siehste}
\ees
where $n_{gh}$ denotes the momentum of the ghost state. We see that the
contribution of the state in $\Hz{0}$ to the eigenvalue is decreased by
applying $G_a^{-n}$ to it. We also note that the new state $G_a^{-n}\hws{k}$ is
in a different representation of $SU(2)$. The new representation is given by
the
tensor product
\be
{\bf k} \times {\bf 1} = {\bf (k + 1)} + {\bf k} + {\bf (k - 1)}
\ee
Since we have seen that applying $G_a^{-n}$ on a state decreases its eigenvalue
for $L_0$ we conclude that only states which are in the representation
${\bf (k-1)}$ can be solutions. According to our discussion given before
we infer that there are solution at ghost level $k-1$ of the form
\be
\Delta_+(G^{-n}_a\hws{k}\hws{k-1,k-1}_{gh})_0\; = 0 \;\; ,
\ee
if $n$ satisfies
\be
n\leq k\;\; .
\ee
There are also solutions at ghost level $k$ of the form
\be
\Delta_+(G^{-n}_a\hws{k}\hws{k,k-1}_{gh})_0\; = 0 \;\; ,  \label{mineen}
\ee
if $n$ satisfies
\be
n\leq k - 1\;\; .
\ee
In order to determine all ghost levels which allow for solutions containing
the state $G^{-n}_a\hws{k}\;$ we note that the minimal momentum of a ghost
state $\hws{l,k-1}_{gh}\;$ at ghost level $l$ in the representation
${\bf (k-1)}$ is
\be
D_0\hws{l,k-1}_{gh}\; \leq -\left({1\over 2}l(l+1) + (k-1)(k-l-1)\right)
\hws{l,k-1}\;\;\; ,\;\; {2\over3}l\leq (k-1)\leq l-1\;\; . \label{mimo1}
\ee
With this relation we conclude that there are solutions at ghost level $l$
of the form
\be
\Delta_+(G^{-n}_a\hws{k}\hws{l,k-1}_{gh})_0\; = 0 \label{solala}
\ee
if the conditions
\be
n\leq 2k -l -1-{1\over 2}(l-k)(l-k-1)\;\; \wedge\;\; n\geq 1
\label{beding}
\ee
can be satisfied.

For states of the form
$G^{-n_1}_{a_1}G^{-n_2}_{a_2}\hws{k}\;$
 there are also
solutions of the form  (\ref{solala}) but the conditions  on the Fourier
indices of the Kac-Moody generators are now
\be
n\leq 2k -l-1-(l-k)(l-k-1)\;\; \wedge\;\; n_1\geq 1\;\wedge\; n_2\geq 1\;\; .
\label{beding1}
\ee
However there are now also states of the form
 $G^{-n_1}_{a_1}G^{-n_2}_{a_2}\hws{k}\;$ which are in the ${\bf (k-2)}$
representation. For those states there are new solutions with ghosts in the
representation ${\bf (k-2)}$. They can be constructed in a completely analogous
way.

In general we can construct solutions for any state $\ket$ generated from
a highest weight state $\hws{k}\;$ by determining its  eigenvalue for $L_0$,
which will be its contribution to the eigenvalue of an eigenstate for
$\Delta_+$ as in (\ref{siehste}). Furthermore one has to find out in which
irreducible representation of $SU(2)$ it falls. With this information one
can construct the corresponding ghost state in the same representation at
a given ghost level. With the knowledge of the minimum momentum of this
ghost state one can easily derive conditions like (\ref{beding}) and
(\ref{beding1}) which are necessary and sufficient for a solution to exist.

So far we have only given in (\ref{mimo1}) the minimum momentum of a ghost
state $\hws{l,k}_{gh}\;$ for the range ${2\over 3}l\leq k\leq l$. We now list
the minimal momenta for the remaining possibilities. For this we have to
distinguish between the cases where ${\bf k}$ is even and the case where
${\bf k}$ is odd. It is also convenient to rewrite $l$ as
\be
l=3l_1 + r\;\; ,
\ee
with
\be
l_1=int(l/3) \; ,\; r=l-3l_1\;\; .
\ee
With this definition the minimum momenta are
\begin{itemize}
\item[1. ] $ k=2k_1$ and $1\leq k\leq {2\over 3}l$:
\be
D_0\hws{l,k}_{gh}\; \geq -\left({3\over 2}l_1(l_1+1) + k_1^2 + r(l_1 +1)\right)
\hws{l,k}_{gh}\;\; ;
\ee
\item[2. ] $ k=2k-1$ and $1\leq k \leq {2\over 3}l$:
\bes
D_0\hws{l,k}_{gh}\; \geq \left\{ \begin{array}{cc}
-\left({3\over 2}l_1(l_1+1) +k_1(k_1+1) +1\right)\hws{l,k}_{gh} & r=0\; ,\\
 & \\
-\left({3\over 2}l_1(l_1+1) +k_1(k_1+1) +r(l_1+1)\right)\hws{l,k}_{gh}& r\neq 0
\; ;\end{array}\right.
\ees
\item[3. ] $k=0$:
\bes
D_0\hws{l,0}_{gh}\;\geq -\left\{ \begin{array}{lr}
-{3\over2}l_1(l_1+1)\hws{l,0}_{gh}\; & r=0\;\; ,\\  \\
-\left({3\over2}l_1(l_1+1)+l_1 +2\right)\hws{l,0}_{gh}\; & r=1\;\; ,\\  \\
-\left({3\over2}l_1(l_1+1)+2l_1 +3\right)\hws{l,0}_{gh}\; & r=2\;\; .
\end{array}\right.
\ees
\end{itemize}
With these informations it is possible to construct all solutions for a given
state in a highest weight representation along the lines of our discussion
given above.

{}From this we see that at ghost level $k > 1$ all representations ${\bf l}$ of
the finite dimensional $SU(2)$ with $l\leq k$ appear as zero modes of the
BRST Laplacian. However, the inner product on these higher ghost levels is
indefinite. To show this we construct out of a solution at ghost level $k$
a new solution at the same ghost level which has a norm with a sign opposite
to the norm of the first one. To do this it is convenient to rewrite the
$SU(2)$
Kac-Moody algebra in terms of step operators:
\be \begin{array}{rcl}
[E^m_+ , E^n_- ] &=& E_3^{m+n} - 2 m \delta^{m,-n}\;\; ,\\
 & & \\
{}[E_3^m , E_3^n ] &=& -2 m \delta^{m,-n}\;\; ,  \\
 & & \\
{}[E_3^m , E^n_{\pm}] &=& \pm E^{m+n}_{\pm}\;\; . \end{array}
\ee
As the first solution we take the highest weight state $\hws{k}$ with $k>1$.
As we have seen this state appears as a solution at the $k$th ghost level.
Since the inner product on the ghost states is positiv definite we only need
to consider the inner product of the YM-states. We set the norm of $\hws{k}$
to
\be
\|\hws{k}\| = \mu \neq 0 \;\; .
\ee
According to (\ref{mineen}) we can construct a new solution out of $\hws{k}$
at ghost level $k$ by applying a step operator on this state with condition
that the new state is in the rep. ${\bf k-1}$ of $SU(2)$. Such a state is
given by
\be
\ket =
\left( E^{-n}_- -{1\over k} E_3^{-n}E^0_- - {1\over k(2k-1)}
E^{-n}_+E_-^0E_-^0\right)
\hws{k}\;\; ,
\ee
where n has to fulfill
$$ n\leq k-1 \;\; .$$
We may check that $\ket$ is a highest weight state of the finite dimensional
$SU(2)$ in the rep. ${\bf k-1}$, i.e.
\be
E^0_+\ket = 0 \;\; \wedge \;\; E_3^0\ket = (k-1)\ket\;\; .
\ee
The norm of this state is given by
\be \begin{array}{rl}
\|\ket\|^2 &= <\!\! {\bf k}|\left( E^n_+ -{1\over k} E^0_+E^n_3 -
{1\over k(2k-1)}E^0_+E^0_+E^n_-\right)
\left( E^{-n}_- -{1\over k} E_3^{-n}E^0_- - {1\over k(2k-1)}
E^{-n}_+E_-^0E_-^0\right)\hws{k} \\
 &= \left(k -1 - {1\over k} - 2n(1 + {1\over k} + {1\over k(2k-1)} \right)
\mu^2\;\; .
\end{array}\label{negativ}
\ee
Setting $n = k-1$ in (\ref{negativ}) we obtain
\be
\|\ket \|^2 = \left( {5k-1\over k(2k-1)} -k - 1 \right) \mu^2
\ee
from which we conclude that the norms of $\ket$ and $\hws{k}$ have opposite
signs for $k>1$. Thus we have shown that the inner product on the cohomology
classes at higher ghost levels is indefinite.

\section{Conclusions}
In this article we have studied 2-d YM-theory on a circle which has an
anomaly already for the pure YM-system, when quantized in an oscillator
basis. The constraints generate a Kac-Moody algebra with negative central
charge for a pure YM-Mills system, which becomes positive when chiral fermions
in a sufficiently large representation of the gauge group are coupled to the
system.

We have performed a BRST quantization for that system and constructed a
non-Hermitian
BRST operator which is nilpotent for any central charge. We used this BRST
operator to define the BRST-Laplacian and used it to study the cohomology of
the BRST operator. We showed that a split of the BRST operator into a part
which corresponds to the first-class constraints and  part corresponding to
the second-class constraints can be used to fix a gauge of the system. We
noted that the BRST operator for the second-class constraints on the gauge
fixed state space is the ``square root'' of the energy momentum operator.

Furthermore it was shown that the cohomology of the non-Hermitian BRST
operator for a Kac-Moody algebra with central charge $\alpha_T >-\aYM$ is
equivalent to the cohomology of the corresponding finite-dimensional
Lie-algebra.
For Kac-Moody algebras with $\alpha_T\leq -\aYM$ we found an infinite set of
solutions which were discussed in some detail for a $SU(2)$-Kac-Moody
algebra. However, the inner product on these cohomology classes turned to be
indefinite.

{\dickmi Acknowledgments}

It is a pleasure to thank J.W.~van Holten and M.~Blau for many helpful
discussions and comments.

\bye